# Shape Changes of Liquid Crystal Elastomers Swollen by Low Molecular Weight Liquid Crystal Drops

*Mahesha Kodithuwakku Arachchige, Rohan Dharmarathna, Paul Fleischer, Antal Jákli\**

M. Kodithuwakku Arachchige, P. Fleischer, A. Jákli
Department of Physics, Kent State University, Kent, OH 44242, USA
E-mail: ajakli@kent.edu

R. Dharmarathna, A. Jákli
Advanced Materials and Liquid Crystal Institute, Kent State University, Kent, OH 44242, USA

An elastomer swelling actuator deforms by absorbing a fluid, thus generating mechanical movement. We show that depositing small droplets of low molecular weight liquid crystal on liquid crystal elastomer (LCE) films leads to shape changes and bending actuation. It is found that the radially symmetric LCE director alignments provide radially symmetric hat shapes, while swelling LCEs with uniform director structure leads to arch shapes. Hybrid samples (different director alignments on two sides) lead to more complicated bent shapes. All the observed shapes can be explained by the diffusion that mainly progresses along the direction normal to the director of the LCE. The swelling induced bending force is elevating the top of the swollen LCE up to a factor of 30, providing a powerful and long-lasting actuation. These observations may lead to applications in various fields, like sealants, soft robotics and biomedical devices.

1. Introduction

An elastomer swelling actuator works by absorbing an organic fluid like an oil, into its polymer network, causing it to swell and deform, thus generating mechanical movement.[1] This process is primarily driven by diffusion, where the fluid molecules move into the elastomer's structure, leading to volumetric expansion and shape change.

Liquid crystal elastomers (LCEs) are lightly crosslinked networks of liquid crystalline polymers. Their existence was predicted by de Gennes in 1975,[2] and they were first synthesized by Finkelmann in 1981.[3] LCEs exhibit many entirely new effects that are not simply enhancements of native liquid crystals or polymers.[4] They combine elasticity of polymers with orientational order of liquid crystals. By contrast to simple polymers, which change shape only in response to external forces, liquid crystal polymers do so spontaneously when they orientationally order their mesogenic segments.[4] Shape changes of LCEs were first observed during heating from the nematic liquid crystal phase to the isotropic state.[5] Since then, various and complex mechanical deformations of LCEs have been achieved by thermal stimuli,[6–12] light,[13–20] humidity,[21] chemicals,[21] electric[22–26] and magnetic fields.[27] More complex shape changes such as bending, twisting, wrinkling etc. have been achieved by introducing patterned director profiles,[13,28] cross-link gradient,[29] patterned crosslinking under photomasks,[30] bilayers, activated surfaces[31] or ions[26] into the LCEs.

LCEs swell equally in three dimensions when both the LCE and the solvent are in their isotropic state.[32] In those systems swelling anisotropy can be induced only by special treatment, such as creating microchannels in isotropic PDMS elastomers by swelling with silicone oil.[33] In their anisotropic liquid crystal phase, LCEs swelling actuation is anisotropic even when the solvent is isotropic, such as found in hydrogen-bonded liquid crystal polymer actuators which show anisotropic swelling in high pH solutions.[34,35] The anisotropic swelling is likely related to the several times larger elastic modulus of a monodomain LCE film along the nematic director than perpendicular to it.[31] Additionally, small angle X-ray diffraction of nematic LCEs indicated certain degree of smectic layering normal to the director,[5] which further restricts diffusion normal to the layers, as it requires permeation. The swelling of LCEs in liquid crystalline solvents has been studied for more than a decade.[32,36,37] When swollen in low molecular weight liquid crystals (LMWLCs), monodomain LCEs undergo anisotropic swelling where only the LCE dimensions perpendicular to the director expanded while polydomain LCEs undergo isotropic swelling in all directions.[36] In all these studies, rectangular LCE samples were embedded in LMWLC solvents and swelled uniformly in a plane.

Here we were interested to find out if LMWLC can induce controlled various shape changes in LCE instead of swelling them uniformly, thus providing a variety of actuation shapes. As a first step toward this goal, we investigate the swelling of disc-shape LCE samples by small LMWLC drops placed in the center of the LCE. In addition to studying how the shape evolves in time, we also investigated how different alignments of the LCE samples influence the final shape of the LCE. The results of these simple swelling geometries provide us important clues about how to achieve any kind of complex patterns by programmed ink-jet printing.

## 2. Results and Discussion

The Swelling of various LCE films were tested by water; mineral oils; ionic liquids 1-Butyl-3-methylimidazolium dicyanamide (Molar mass – 205.26 g), 1-Ethyl-3-methylimidazolium ethyl sulfate (Molar mass – 236.29 g), 1-Butyl-3- methylimidazolium hexafluoro phosphate (Molar mass – 284.18 g) and 1-Hexyl-3-methylimidazolium bis (trifluoro methyl sulfonyl) imide (Molar mass – 447.42 g); and the low molecular weight liquid crystal pentyl cyano biphenyl (5CB) purchased from Sigma-Aldrich (Figure S1 and S2 in Supporting Information (SI)). Among those only 5CB was diffusing into the LCE.

After verifying that, the differently aligned 100 μm thick LCE films were cut into 6 mm diameter circular discs with mass of $3.2 \pm 0.1 mg$, and 0.5 μl ($\approx 0.5\ mg$) of 5CB was dropped to the center of the LCE disc as illustrated in **Figure 1**c. Two cameras were set perpendicular to each other to track the swelling of the LCE disk from two orthogonal side views.

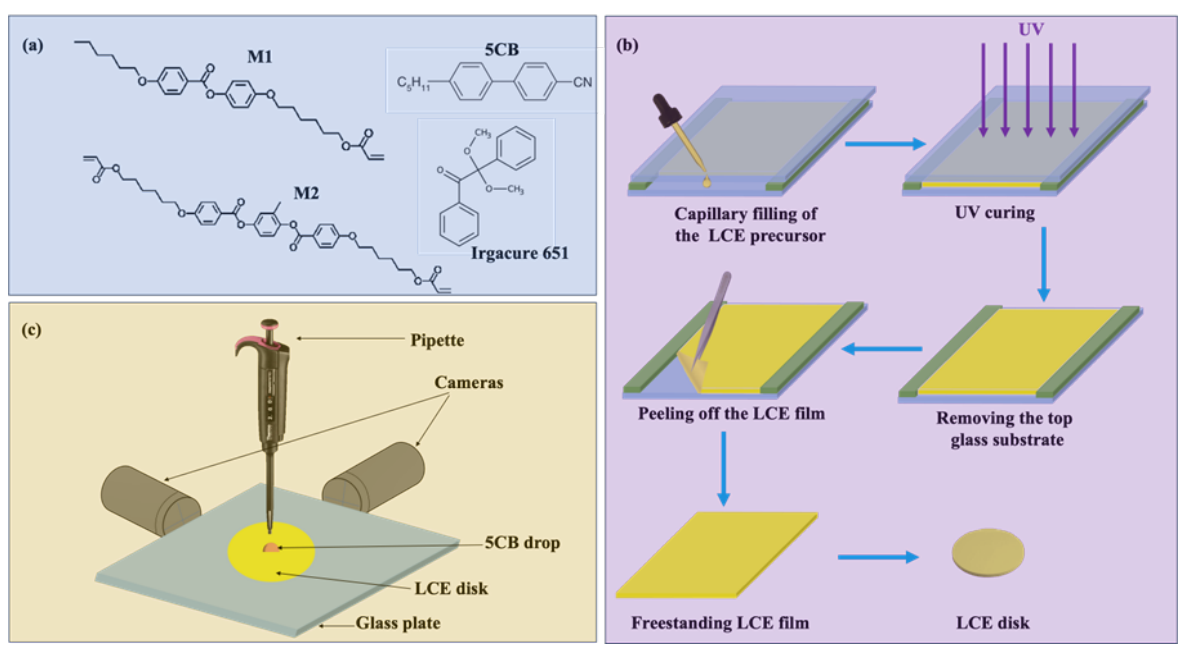

**Figure 1.** Materials used, the LCE preparation steps and the experimental setup. (a) Molecular structures of monofunctional acrylate monomer M1, bifunctional crosslinker M2, photo-initiator Irgacure 651, and low molecular weight liquid crystal 5CB. (b) The steps of fabricating LCE film. (c) Experimental setup of swelling and monitoring LCE disks with 5CB.

X-ray scattering results revealing the nanostructure of a planar aligned LCE are summarized in **Figure 2**a. The inset shows the 2D scattering pattern containing 2 narrower and 1 broad peak centered along the rubbing direction (director). The wave vector ($Q$) dependence of the azimuthally integrated intensity is plotted in the main pane of Figure 2a. The smaller narrow peak is at $0.147 Å^{-1}$, the larger one is at $0.214 Å^{-1}$. They correspond to $42.7 Å$ and $29.4 Å$ spatial periodicities that with good approximation are equal to the stretched molecular lengths of M2 and M1, respectively.

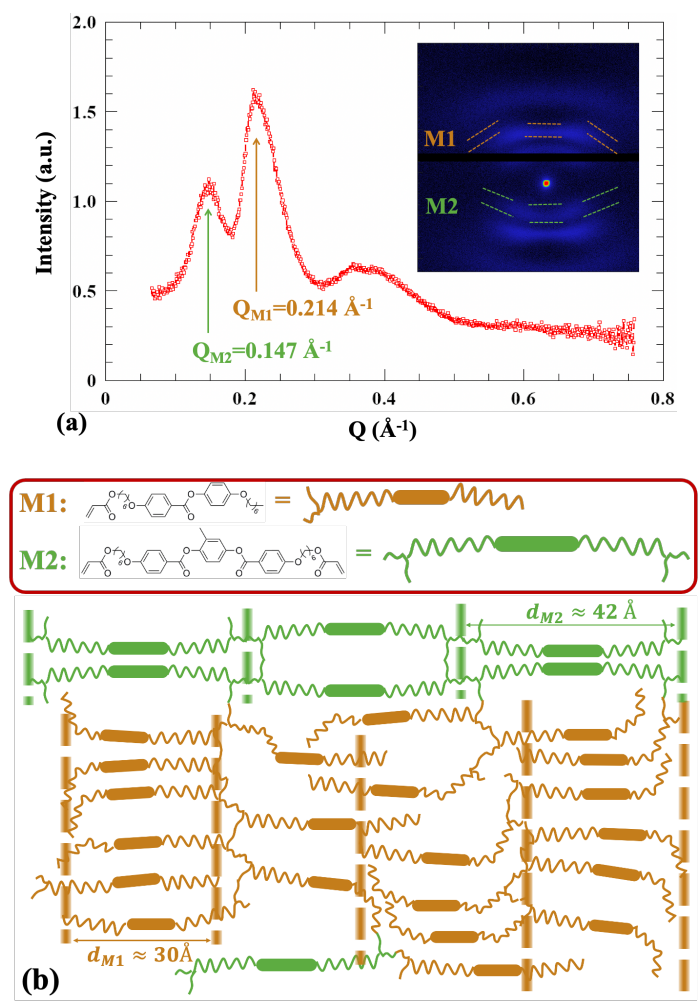

**Figure 2.** Small angle X-ray scattering results and the proposed nanostructure of the studied LCE. (a) Main pane: wave vector (Q) dependence of the azimuthally integrated intensity. Inset: 2D scattering pattern with overlayed illustration of the smectic clusters. (b) Proposed nanostructure of chemically bonded M1 and M2 monomers.

From the width of the peak at Full Width at Half Maxima ($\Delta Q_{FWHM}$) one can obtain the correlation length of the periodicities as $\xi_{M1} \approx \xi_{M2} = \frac{2\pi}{\Delta Q_{FWHM}} \approx 11\ nm$. This means that the average smectic clusters contain about 3-4 layers. The third, broadest and smallest peak can be attributed to be the superposition of the tail of the main M1 peak and of the first harmonics of the M2 and M1 peaks. Such a structure can be interpreted as a result of smectic cybotactic clusters formed by the M1 and M2 monomers, as schematically illustrated in Figure 2b. The 2D pattern shown in the inset of Figure 2a reveals peanut shaped peaks indicative of smectic clusters with uniform director orientation but with a distribution of differently tilted layers,[38] as schematically overlayed on the 2D image. Similar small-angle peaks are observed in isotropic samples; however, the 2D pattern is circular with uniform intensity across all angles, indicating that the smectic clusters are randomly oriented in all directions (Figure S3 in SI). Note that no small-angle peaks are observable in homeotropic cells as there the layers of the smectic clusters are parallel to the substrates, i.e., the periodicity is along the X-ray beam, not providing any Bragg reflection.

Polarizing optical microscopy (POM) images of differently aligned LCE textures at room temperature are shown in **Figure 3**.

The texture of the LCE crosslinked in the isotropic phase appears dark (see Figure 3a) indicating optical isotropy, i.e., random orientation of the director in sub-optical wavelength range. The texture of the LCE with surface treatment that aligns the director normal to the substrate (homeotropic alignment) also appears dark, as shown in Figure 3b. Figure 3c and 3d show the POM textures of an LCE film aligned with substrates treated by unidirectionally rubbed PI 2555 that aligns the director parallel to the substrates (uniform planar alignment). When the rubbing direction of the film is at $\pm 45\,°$ with the crossed polarizers (Figure 3c), the texture is uniformly bright, and when the rubbing direction of the film is parallel to the analyzer or the polarizer, the texture is dark (Figure 3d) proving the alignment is indeed uniformly parallel to the substrates.

The POM texture of the circular planar sample (Figure 3e) shows four dark brushes and a point defect in the center (a.k.a., Maltese Cross) which is consistent with a tangential director alignment we expect for circular rubbing. Figure 3 (f-h) show the POM textures of hybrid aligned cells (one substrate has uniform planar, the other has homeotropic alignment) with crossed (Figure 3f) and oppositely uncrossed (Figure 3g and 3h) polarizers. One can see two domains that appear similar under crossed polarizers, and one is dark while the other is bright under uncrossed polarizers with exchanging brightness under opposite uncrossing.

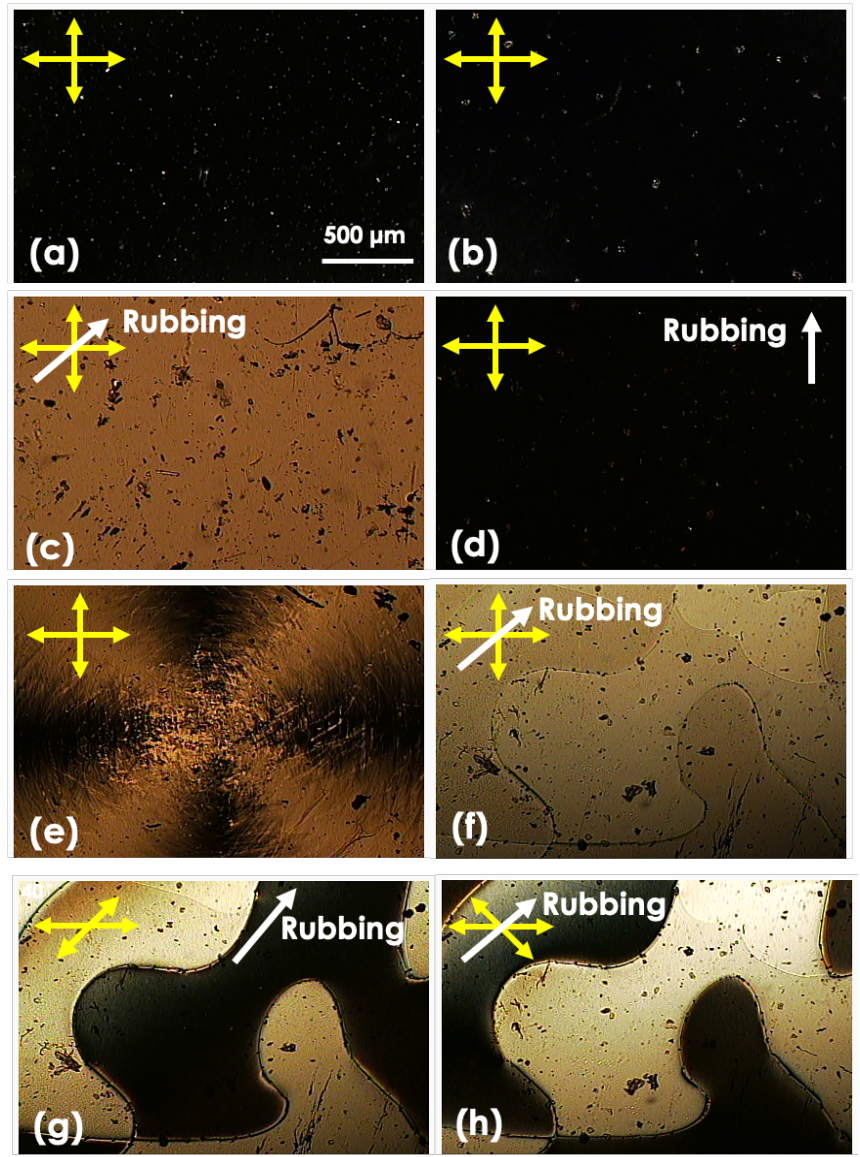

**Figure 3**. POM textures of differently aligned LCEs. (a): LCE crosslinked in the isotropic phase; (b): Homeotropically aligned LCE; (c and d): Uniaxial planar LCE between crossed polarizers with rubbing direction at 45° (c) and 0° (d) to the analyzer; (e): LCE with circular planar alignment; (f-h): Homeotropic-planar hybrid LCE with planar side on top between crossed polarizers and rubbing direction at 45° to the analyzer (f), with analyzer rotated by 45° clockwise (g), and the analyzer rotated by 45° counterclockwise (h).

The shape changes during swelling of isotropic and homeotropic LCE samples by 5CB are shown in **Figure 4** at 0, 0.5, 24 and 264 hours after deposition.

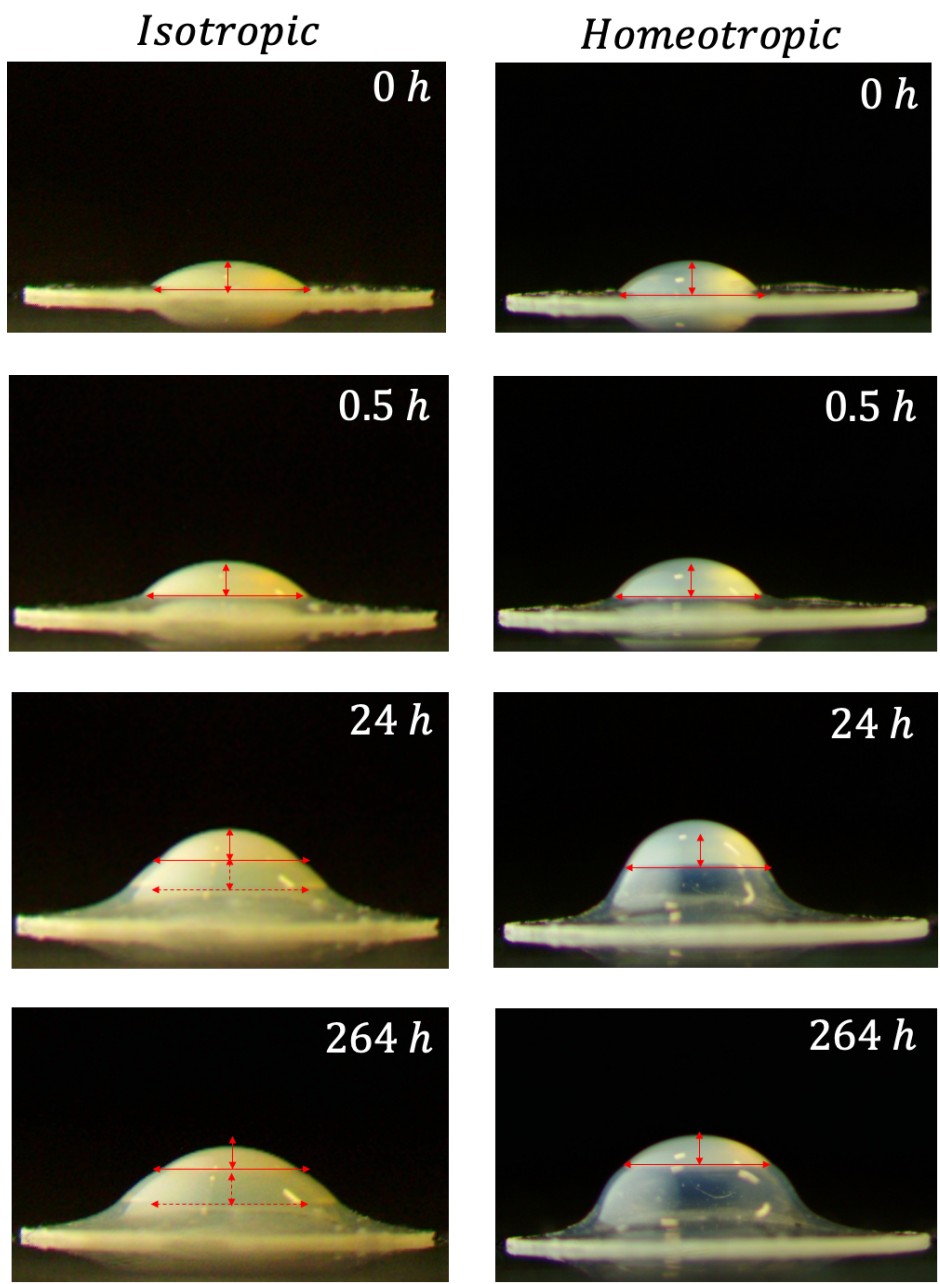

**Figure 4.** Snapshots of swollen LCE samples at 0, 0.5, 24 and 264 hours after deposition of 0.5 mg 5CB drop at the center of a 6 mm diameter 0.1 mm thick LCE discs. Left column: LCE sample crosslinked in the isotropic phase; Right column: Homeotropic LCE sample.

The swelling is radially symmetric in both cases, but the shape evolution is different in the isotropic (left column) and homeotropic (right column) alignments. Note the appearance of the mirror image of the LCE below the glass plate holding the swollen LCE samples. Comparing the original 5CB drops, we can see that the contact angle of 5CB on the isotropic LCE is slightly smaller than on the homeotropic LCE showing that the surface tension of the 5CB and LCE depend on the alignment. To ease the comparison of the shapes at different times of the same alignments, we show the height and width of the original 5CB drop by red arrows and copied those on the pictures taken at different times of the same alignment, as well. In addition to the shape change, one can also follow the progress of the swelling from the change of opacity of the areas, as the 5CB drop and the areas containing 5CB are opaquer than of the pure LCE film that appears to be translucent.

In the swollen isotropic LCE sample $0.5\ h$ after the 5CB deposition, the opaque area appears about 15-20% taller and slightly wider indicating the beginning of the swelling. Additionally, a meniscus area connecting the swollen and original LCE areas appears. After $24\ h$, there are 4 regions observable with different opacities. The cap containing mainly the 5CB drop has about 10% larger height and width than of the original 5CB drop. A secondary opaque range is extending sidewise till the LCE-air interface with width of ~50% and height of ~90% larger than of the original 5CB drop. A third, deeper but slightly narrower, less opaque range appears in the meniscus area that connects to the pure not-swollen LCE film (4th region). As seen in the right column of Figure 4, the homeotropic swollen LCE sample can be divided into areas with only two distinct opacities: one is almost as opaque as the original 5CB droplet, and the other one is almost as transparent as of the pure LCE. After $0.5\ h$, the width of the opaque spherical cap is practically unchanged compared to that of the original 5CB droplet, while its height has increased by about 20%, indicating the start of the swelling. This is also evidenced by the appearance of the meniscus area below the opaque cap. After $24\ h$, the cap width remains practically the same, while its height increases by about 30%. At the same time, the height of the meniscus area increases from $0.1\ mm$ to $0.9\ mm$. After $264\ h$, the opaque cap height decreases back to the original level without appreciable change in its width and, the height of the translucent meniscus area increases by an additional $0.1\ mm$. At this stage, the width of the meniscus area becomes about 30% larger than after $24\ h$.

Comparing the shapes of the homeotropic and isotropic swollen LCEs, we can see that the homeotropic LCE has a "Sombrero" shape, whereas the shape of the isotropic LCE rather resembles to a "Pith-helmet".

As the pictures shown in Figure 4 do not reveal the shape of the inside wall, in **Figure 5**a we show pictures both the outside and inside walls by turning upside-down an isotropic LCE sample 264 h after swollen by the 5CB drop. Comparing the original and upside-down shapes, one can see that the thickness of the swollen sample is basically constant.

The time dependences of the height $h(t)$ of the tip of the swollen LCEs with respect to the height of the original LCE film are shown in Figure 5b. They can be fitted by single exponential functions $h(t) = h(\infty)\left(1 - e^{-\frac{t}{\tau}}\right)$. The parameters of the best fits are $h(\infty) \approx 1.2\ mm$ and $\tau \approx$

5.7 $h$ for the isotropic (red curve) sample, and $h(\infty) \approx 1.5\ mm$ and $\tau \approx 4.3\ h$ for the homeotropic (blue line) sample.

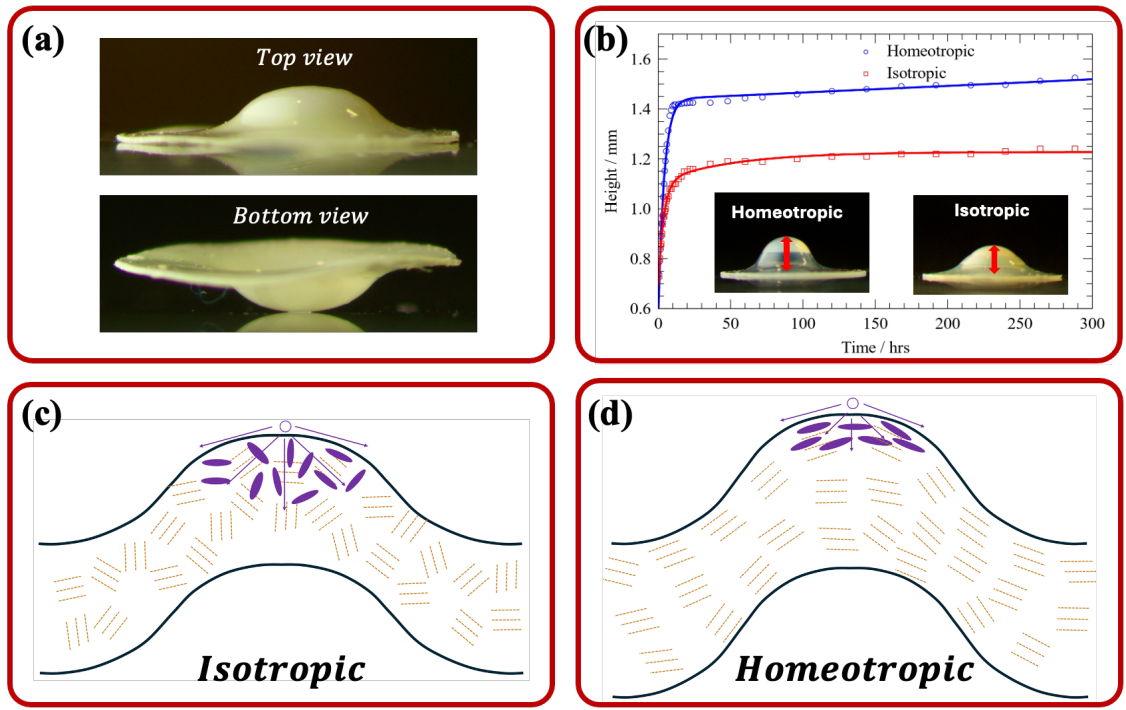

**Figure 5.** Analyses of the shape changes of isotropic and homeotropic LCE samples swollen by a 5CB drop. (a) Side views of an isotropic elastomer with swollen 5CB on top and after turned upside down revealing the inside shape. (b) Time dependence of the height for homeotropic and isotropic samples. (c and d) Cartoons of the diffusion of the 5CB molecules (purple ellipses) into LCE polymerized in the isotropic phase (c) and in the nematic phase between homeotropic substrates (d).

The difference between the shapes of the homeotropic and isotropic LCEs can be understood by the different diffusion processes. In the isotropic LCE the 5CB molecules diffuse downward and sideway with the same intensity (see cartoon in Figure 5c). That is why the opaque area is progressing downward much more than in the homeotropic LCE, where the diffusion is mainly progressing sidewise (see cartoon in Figure 5d), and the opaque cap is distinct from the transparent pure LCE. At the beginning of the swelling, the homeotropic LCE can be modeled as a dilated outer layer that leads to the bending of the 5CB drop area. That strong vertical 5CB density gradient provides vertical stress, thus effectively pulling up the neighboring pure LCE causing the "sombrero" shape. In the isotropic LCE that upward pulling and sidewise pushing are both present as the horizontal gradient of the 5CB molecules lead to a lateral force, which results in lower and wider bulge.

Snapshots of the shapes of planar aligned swollen LCEs are shown in **Figure 6**. The left column shows a sample with circularly rubbed surfaces at 0, 0.5, 24 and 48 $h$ after deposition of the 5CB drop. One can see that, similar to the isotropic and homeotropic samples shown in Figure 4, the shape has radial symmetry. The difference is that the elevation is smaller, and there is a peak in the middle where there is a defect of the director, as seen in Figure 3e. Such a peak is consistent with prior observations on other LCE films with defect patterned surfaces.[39] Note that after long-term the central elevation around the circular defect often pokes creating a hole in the place of the defect as shown in the bottom picture after 48 $h$ of 5CB deposition.

The middle and right columns show a sample with uniformly rubbed planar surfaces with rubbing direction (magenta arrow) along the plane (middle column) and rubbing direction (magenta arrowhead) perpendicular to the plane of view (right column). Unlike the swelling of the isotropic, homeotropic and circularly planar LCEs, the swelling is not radially symmetric for the uniformly planar LCE sample. Notably even half an hour after the deposition of the 5CB drop, the LCE bends in the plane perpendicular to the director. This, and the extended shape of the diffusing 5CB patch normal to the director indicate that the 5CB molecules are preferably diffusing perpendicular to the director as a consequence of the larger elastic modulus of monodomain LCEs along the director. The slightly anisotropic bending shape is due to the small off-central position of the 5CB drop that can be judged from the top-right picture of Figure 6. We note that the bent-shapes of the swollen LCE seen in Figure 6 resemble to that observed in hybrid aligned LCE films under heating.[12,28] In that case it is related to the different thermal expansion coefficients of the LCE at the planar and homeotropic side. Comparing the magnitudes of the bending, similar strains induced the thermal actuation of hybrid films require typically up to 100 °C heating.

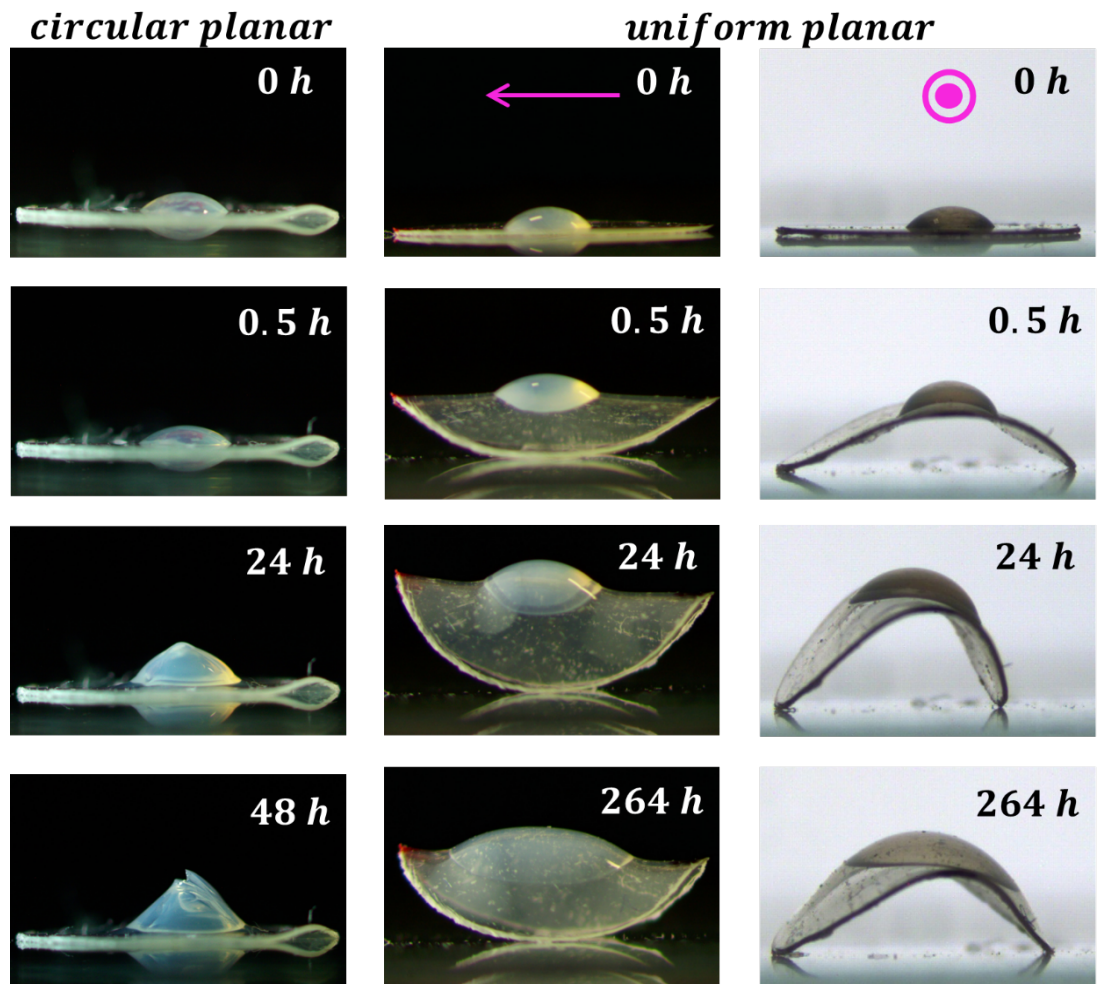

**Figure 6.** Snapshots of swollen planar LCE samples at different times after deposition of 0.5 mg 5CB drop at the center of a 6 mm diameter 0.1 mm thick LCE discs. Left column: Circularly rubbed planar surfaces; Middle and right columns: uniformly rubbed planar surfaces. Middle column: Rubbing direction (magenta arrow) along the plane; Right column: Rubbing direction (magenta arrowhead) perpendicular to the plane of view.

Analyses of the shapes of the uniform planar samples are shown in **Figure 7**. Figure 7a and 7b show the time dependences of the heights as measured with the red lines shown in the inset pictures. The time dependence of the height from the tip to the top of the sample shows a sharp decrease in the first few hours indicating the swelling of the 5CB drop begins almost instantaneously. The height reaches a minimum of about 0.32 $mm$ after ~ 5 h, then increases to reach ~0.47 $mm$ after ~30 $h$. After that, the height remains basically constant in the studied 200 $h$ time interval. The time dependence of the height compared to the plate the sample is placed on, is shown in Figure 7b. The height increases sharply in the first few hours reaching 2.9 $mm$ after ~8 $h$. Then $h(t)$ shows a damped oscillation with a period of about $20 - 25\ h$, revealing the visco-elastic nature of the bent LCE disc. Additionally, one can see that on long-term the height decreases. This is related to the slowly progressing diffusion that eventually leads to a uniform swelling of the entire LCE disc. Once it happens, the disc becomes uniformly flat again as the swelling stress is uniform. In fact, after 9 months of swelling (Figure S4 in SI), the planar sample becomes completely flat. Note that during the same time interval, the isotropic and the homeotropic LCE films still had some curvature (more for the homeotropic), further proving that the diffusion is slowest in the homeotropic sample where the diffusion along the director is very slow, since they require permeation through the smectic clusters. We also note that the flattening can be also forced by depositing the same amount of 5CB droplet on the opposite side of the LCE film.

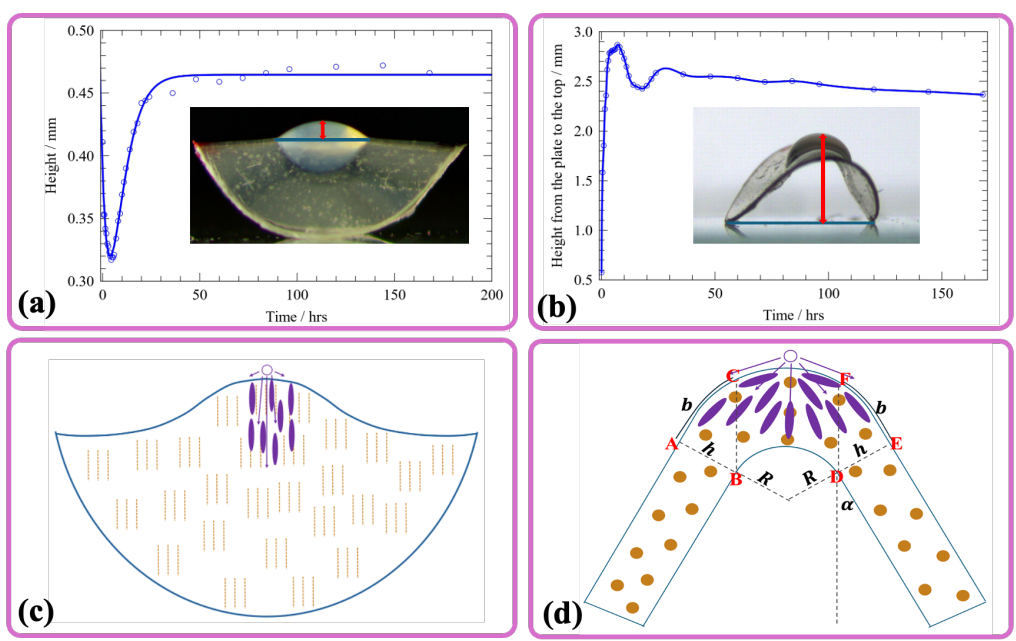

**Figure 7.** Analyses of the shapes of the uniform planar samples. (a and b): The time dependences of the heights as measured with the red lines shown in the inset pictures. (a): Time dependence of the height from the tip to the top of the sample; (b): Time dependence of the height from the tip to the supporting plate; (c and d): Illustration of the shape and diffusion of the 5CB molecules (purple ellipses) along the plane of the LCE sample (c), and perpendicular the plane and the director (d). Orange dots in (d) show the director of LCE is perpendicular to the view.

The swelling induced bending observed in the planar LCE film can be understood as follows. As the diffusion propagates from the top, the top side of the film is stretched compared to the bottom, resulting in a bending stress on the LCE film. As it can be seen in Figure 7b, such a bending stress leads to an elevation of the center of the LCE film from the original $0.1\ mm$ (the thickness of the film) to $2.9\ mm$ after 8 hours of the 5CB deposition. This $\approx 30-fold$ lift is really remarkable, since the weight of the 5CB drop results in a downward gravity force of $F_g = m_{5CB} \cdot g \approx 5\ \mu N$. The bending force leads to an increase of the potential energy in the order of $U < h_{max}g(m_{5CB} + m_{LCE}) \approx 2.9 \cdot 10^{-3} \cdot 10 \cdot (5 \cdot 10^{-4} + 3.2 \cdot 10^{-3}) \cdot 10^{-3} \approx 10^{-7} J$. The bending energy has to overcome this value. An elastic plate with Young's modulus $E$, curvature radius $R$, curved area $A$, thickness $h$ and Poisson's ratio $v$, has a bending energy $W_B = \frac{1}{2} \iint R(x,y)^{-2} \frac{Eh^3}{1(1-v^2)} dxdy \approx \frac{Eh^3 A}{24R^2(1-v^2)}$.[40] Taking $E \approx 10\ MPa$ Young's modulus value for the planar LCE,[26] $R \sim 1mm$ (see Figure 6), $A \approx R^2$, $v \approx 0.4$ and $h \approx 100\ \mu m$, we get $W_B \sim 5 \cdot 10^{-7} J$. This is indeed larger than the bending induced increase of the potential energy, showing that the Young's modulus of the LCE is large enough to sustain the observed mechanical actuation.

To estimate if the volume of the 5CB drop is indeed enough for the observed deformation of the planar sample, we approximate the cross - sectional shape (see the right image of Figure 6 at 24 h) as two straight legs with angle $\alpha$ compared to the vertical direction as shown in Figure 7d. Assuming no change in density of the swollen parts, the volume of the swollen 5CB ($V_{S5CB}$) that provides the increased volume of the bent LCE, can be calculated by the areas surrounded by the points A,B,C and D,E,F multiplied by the width $w$ of the film (see Figure 7d). With $\alpha \approx 45°$, this can be approximated as $V_{5CB} \leq h^2 w \sim 10^{-8} \cdot 6 \cdot 10^{-3} \sim 6 \cdot 10^{-11}\ m^3$. This is smaller than the total volume $V_{5CB}$ of the 5CB drop that is $V_{5CB} \approx 0.5\ \mu l = 5 \cdot 10^{-10}\ m^3$, showing that the deposited 5CB droplet is more than enough to induce the required bending. Finally, we studied swelling of three hybrid samples. The first one (left column of **Figure 8**) has homeotropic alignment at the top and degenerate (non-rubbed) planar alignment at the bottom. As both alignments are radially symmetric, the shape also stayed radially symmetric for a long time. The small asymmetry seen after 60 hours, is due to the slight off-centered position of the original 5CB drop, as seen in the top-left picture of Figure 8.

The sample with homeotropic alignment on the top and uniformly planar at the bottom (see middle column of Figure 8) becomes bent after 12 hours due to the asymmetry of the alignment. Similarly, the sample with uniform planar alignment on the top and homeotropic on the bottom (right column of Figure 8) becomes bent after 12 hours, but surprisingly it becomes inverted when becomes bent, i.e., it is bulging downward instead of upward from where the 5CB was deposited. This is related to the faster diffusion normal to the director, i.e., on the planar side, leading to accumulation of the 5CB molecules at the bottom, on the opposite side of the deposition.

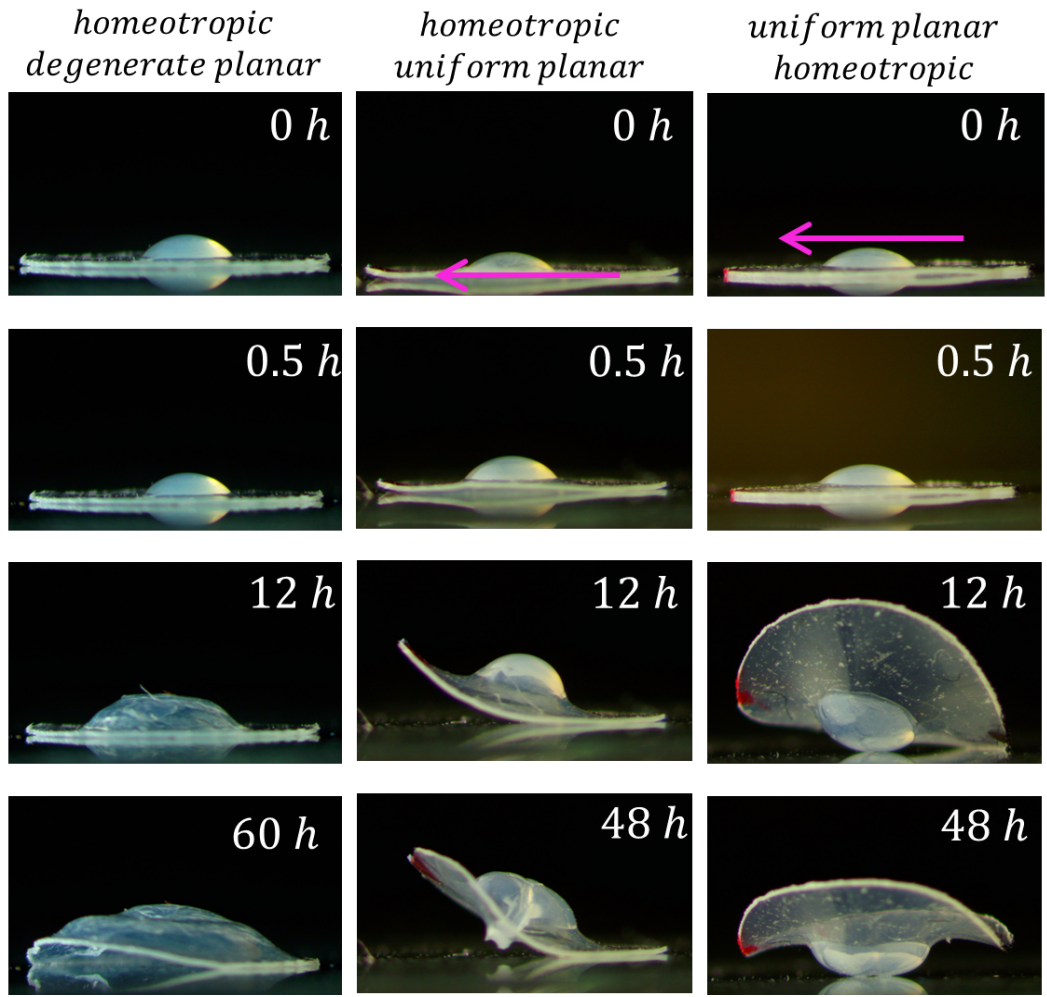

**Figure 8.** Shape changes of swollen hybrid samples. (a) Homeotropic-degenerate planar hybrid LCE sample with homeotropic on top (b) Homeotropic-planar hybrid LCE sample with homeotropic on top. (c) Homeotropic-planar hybrid LCE sample with planar on top.

## 3. Conclusions

We have shown that depositing small droplets of low molecular weight liquid crystal 5CB on liquid crystal elastomer films leads to shape changes and bending actuation of the swollen elastomer. For simplicity, we have studied droplets deposited in the middle of the top of the cylindrically symmetric disc-shaped elastomer samples. It is found that the radially symmetric isotropic, homeotropic and circularly rubbed planar LCE alignments provide radially symmetric shapes. Swelling of uniformly planar aligned samples lead to arch shapes with arch plane being perpendicular to the director. Hybrid samples (one side is homeotropic, the other side is planar) form more complicated bent shapes, including inverted umbrella shapes. All these shapes can be explained by anisotropic diffusion of the 5CB molecules into the LCE, namely that the diffusion is mainly progressing normal to the director of the LCE.

Remarkably, in spite of the downward gravitational force acting on LCE and the deposited 5CB drop, the swelling induced bending force is elevating the top of the swollen LCE up to a factor of 30, providing a very powerful and long-lasting actuation. This can have application in various fields, such as sealants,[41,42] soft robotics where controlled expansion is needed, and biomedical devices. We also envision the creation of complex surface patterns by depositing multiple droplets with varying size and positions.

## 4. Experimental Section/Methods

LCE precursors (Figure *1*a) were made by mixing a monofunctional acrylate monomer M1 (4-(6-Acryloxy-hex-1-yl-oxy) phenyl 4-(hexyloxy) benzoate) and bifunctional crosslinker M2 (1,4-Bis [4-(6 - acryloyloxyhexyloxy) benzoyloxy]-2-methylbenzene) purchased from Synthon chemicals with a photo-initiator (2,2- Dimethoxy-2-phenylacetophenone (Irgacure® 651)) in 87:12:1 weight ratio.[43] A few drops of Dichloromethane were added to the mixture and sonicated for 2-3 minutes and let the solvent evaporate overnight at room temperature. Then it was heated to 90 °C (isotropic temperature of the LCE precursor) and mechanically stirred for 10 minutes to achieve complete mixing.

To prepare LCE samples, sandwich cells with differently aligned glass substrates and 100 μm thick spacers between them were used. Glass substrates were sonicated in detergent (Cavi-clean ultrasonic detergent) water at 60 °C for 15 minutes and then washed with distilled water and isopropyl alcohol (IPA), respectively followed by drying at 90 °C for 15 minutes. The top glass substrate was secured to the cell using two clips. The steps of fabricating a 100 μm thick LCE film are shown in Figure *1*b.

The LCE precursor was filled into the cell by capillary filling at 90 °C and then crosslinked under UV light for 5 minutes. Five types of LCE samples labeled as isotropic, homeotropic, uniaxial planar, circular planar and hybrid were prepared. The isotropic sample was crosslinked at 90 °C in the isotropic phase of the LCE precursor between two clean glass substrates with no alignment layer. To align the mesogenic units parallel (planar alignment) and perpendicular (homeotropic alignment) to the substrates, polyimide PI-2555 and SE-5661 were used, respectively. Polyimides were spin-coated onto the glass substrates and then soft baked at 90 °C for 5 minutes and hard baked at 275 °C for 1 hour. The homeotropic, uniaxial planar, circular planar, homeotropic-planar hybrid and homeotropic-degenerate planar hybrid samples were crosslinked at 55 °C in the nematic phase of the LCE precursor between two glass substrates with homeotropic, uniaxial planar, circular planar alignment layers on both substrates, and uniaxial planar or non-aligned layer on one substrate and homeotropic alignment layer on the other surface. To align PI 2555 unidirectionally, we rubbed it with a block covered with velvet cloth. For circular rubbing of PI2555, we spun the substrate at 30 rpm for 20 seconds under a fixed rubbing block.

**Supporting Information**

Supporting Information is available from the Wiley Online Library or from the author.

**Acknowledgements**

P.S. acknowledges support from the Summer Undergraduate Research Experiences (SURE) program of Kent State University.

**Conflict of Interest Statement**

The authors declare no conflict of interest.

**Data Availability Statement**

The data that support the findings of this study are available from the corresponding author upon reasonable request.

# Supporting Information

**Shape Changes of Liquid Crystal Elastomers Swollen by Low Molecular Weight Liquid Crystal Drops**

*Mahesha Kodithuwakku Arachchige, Rohan Dharmarathna, Paul Fleischer, Antal Jákli\**

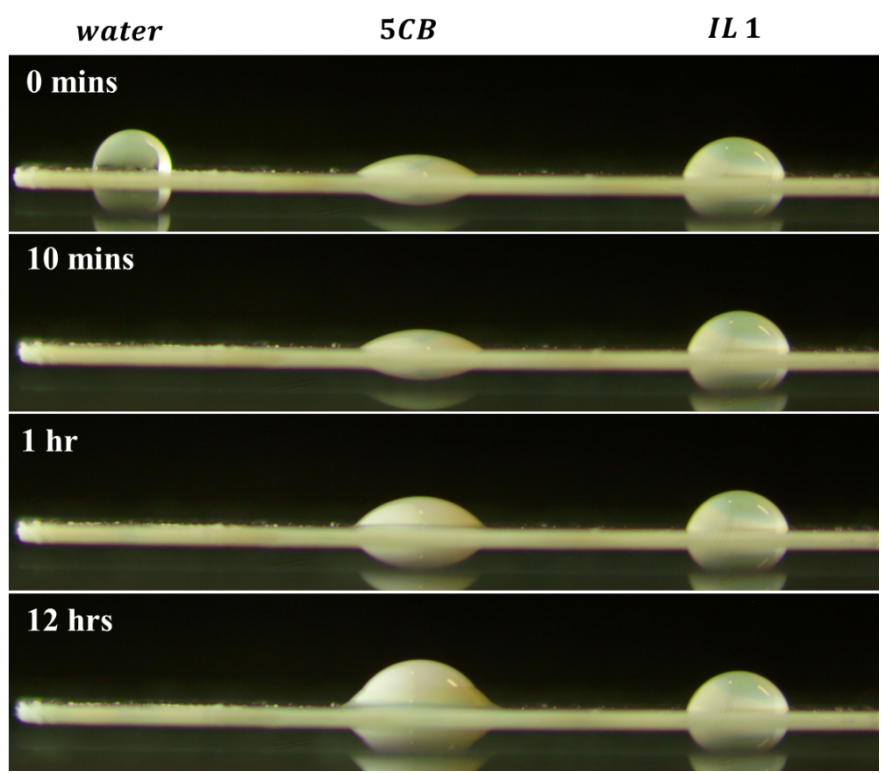

**Figure S1.** A strip of isotropic LCE swollen with 0.5 µl drops of water (left), 5CB (middle) and the ionic liquid, 1-Butyl-3-methylimidazolium dicyanamide (IL 1).

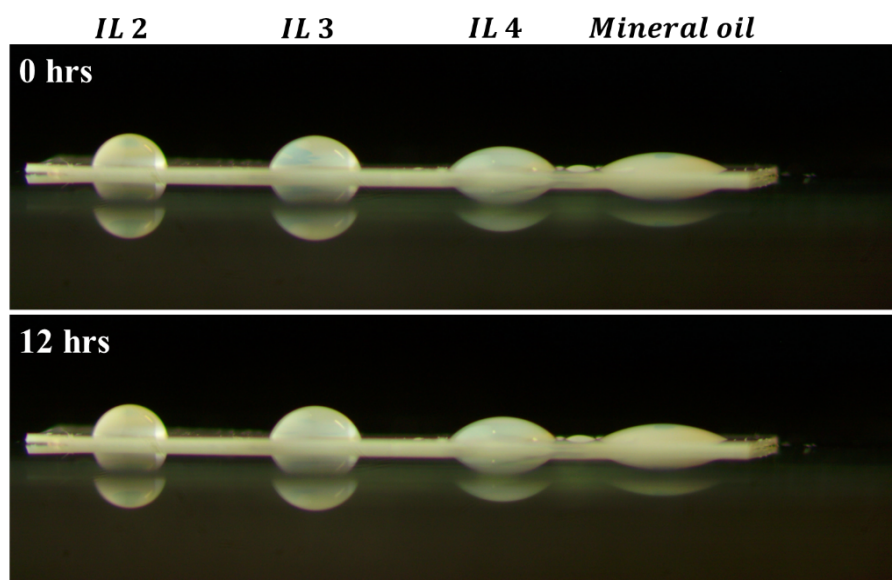

**Figure S2.** A strip of isotropic LCE swollen with 0.5 μl drops of the ionic liquids 1-Ethyl-3-methylimidazolium ethyl sulfate (IL 2), 1-Butyl-3- methylimidazolium hexafluoro phosphate (IL 3) and 1-Hexyl-3-methylimidazolium bis (trifluoro methyl sulfonyl (IL 4), and mineral oil (right).

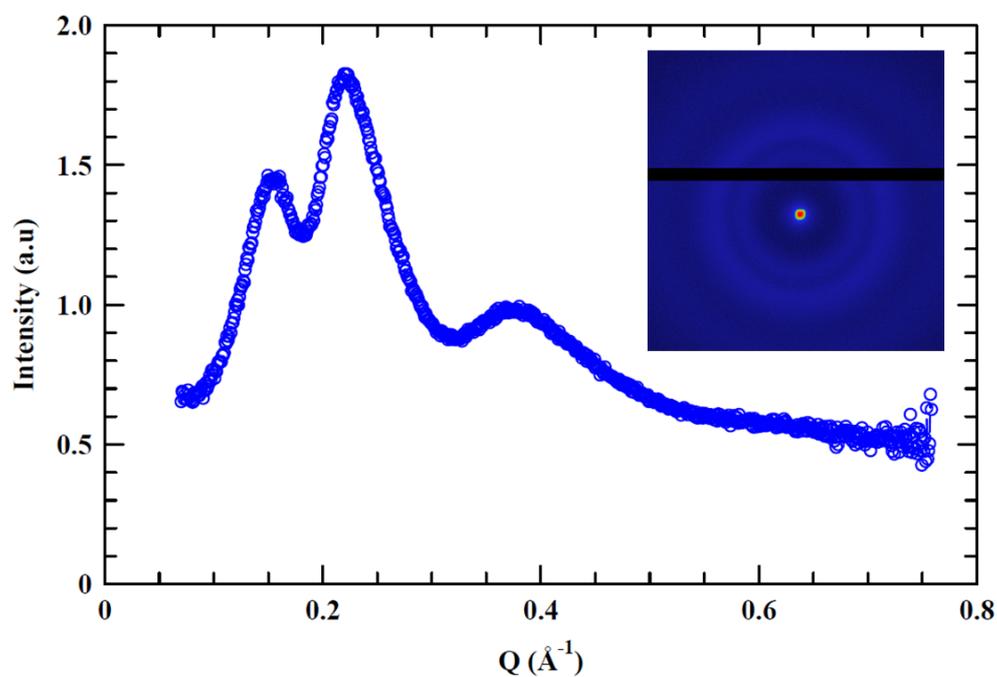

**Figure S3**. Small angle X-ray scattering results for isotropic LCE sample. Main pane: wave vector (Q) dependence of the azimuthally integrated intensity. Inset: 2D scattering pattern.

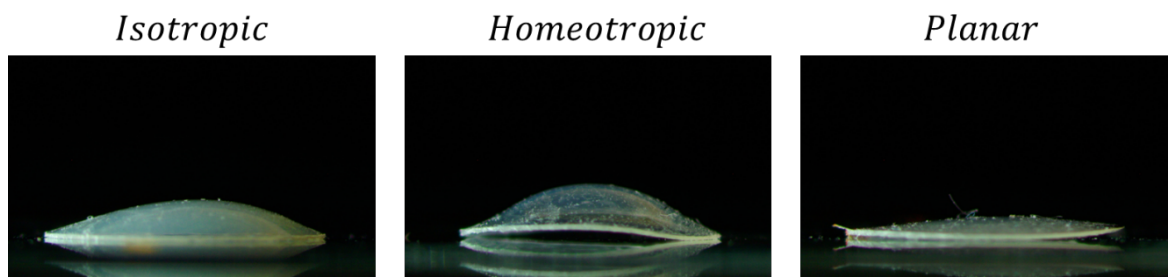

**Figure S 4.** Isotropic, Homeotropic and Planar samples after 9 months of swelling

3WILEY-VCH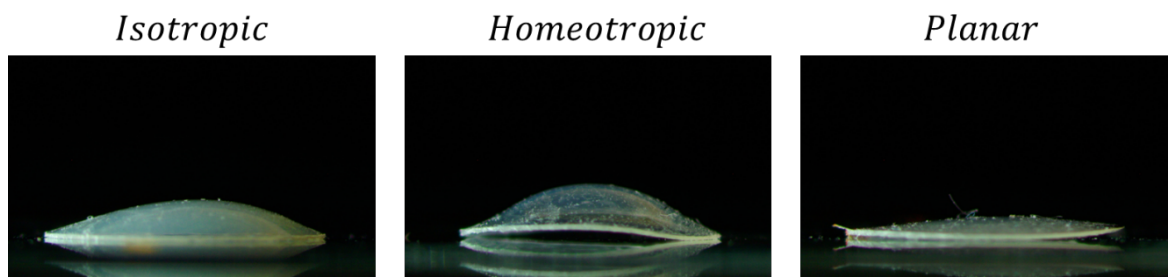

**Figure S 4.** Isotropic, Homeotropic and Planar samples after 9 months of swelling

3